*Research Article*

# Myoelectric Pattern Recognition Performance Enhancement Using Nonlinear Features

Md. Johirul Islam[1,2], Shamim Ahmad[3], Fahmida Haque[4], Mamun Bin Ibne Reaz[4,*], Mohammad A. S. Bhuiyan[5,*], Md. Rezaul Islam[1]

[1]Dept. of Electrical and Electronic Engineering, University of Rajshahi, Rajshahi-6205, Bangladesh
[2]Dept. of Physics, Rajshahi University of Engineering and Technology, Rajshahi-6204, Bangladesh
[3]Dept. of Computer Science and Engineering, University of Rajshahi, Rajshahi-6205, Bangladesh
[4]Dept. of Electrical, Electronic and Systems Engineering, Universiti Kebangsaan Malaysia, 43600 UKM, Bangi, Selangor, Malaysia
[5]Dept. of Electrical and Electronic Engineering, Xiamen University Malaysia, Bandar Sunsuria, 43900 Sepang, Selangor, Malaysia
*Author to whom any correspondence should be addressed.

E-mail: mamun@ukm.edu.my and arifsobhan.bhuiyan@xmu.edu.my

**Keywords:** EMG pattern recognition, feature extraction, LMAV, NSV, nonlinear feature.

## Abstract

The multichannel electrode array used for electromyogram (EMG) pattern recognition provides good performance, but it has a high cost, is computationally expensive, and is inconvenient to wear. Therefore, researchers try to use as few channels as possible while maintaining improved pattern recognition performance. However, minimizing the number of channels affects the performance due to the least separable margin among the movements possessing weak signal strengths. To meet these challenges, two time-domain features based on nonlinear scaling, the log of the mean absolute value (LMAV) and the nonlinear scaled value (NSV), are proposed. In this study, we validate the proposed features on two datasets, existing four feature extraction methods, variable window size and various signal to noise ratios (SNR). In addition, we also propose a feature extraction method where the LMAV and NSV are grouped with the existing 11 time-domain features. The proposed feature extraction method enhances accuracy, sensitivity, specificity, precision, and F1 score by 1.00%, 5.01%, 0.55%, 4.71%, and 5.06% for dataset 1, and 1.18%, 5.90%, 0.66%, 5.63%, and 6.04% for dataset 2, respectively. Therefore, the experimental results strongly suggest the proposed feature extraction method, for taking a step forward with regard to improved myoelectric pattern recognition performance.

# 1. Introduction

From the perspective of performing regular activities after limb loss or from the view of people born with congenital defects, artificial limbs or prostheses are very helpful [1]. Many modern prostheses, such as i-Limb [2], Cyberhand [3], and Yokoi Hand [4], use EMG signals to control multiple degrees of freedom of prosthesis movements since the pattern recognition because it is non-invasive and convenient for long-term data acquisition [11-13]. In addition to EMG signal, force myography (FMG), a non-invasive technique for measuring the pressure patterns between the underlying muscle and the pressure sensor during muscle contraction, is also used for upper limb prosthetic control [14-17]. Also, mechanomyography (MMG) is another alternative technique that measures vibrational characteristics during muscle contraction employing an accelerometer or microphone [18-20]. But the frequency spectrum of EMG is wide compared to FMG and MMG and carries more information corresponding to a muscle contraction [21].

In myoelectric pattern recognition, the features, the vital components of myoelectric pattern recognition, are extracted from the EMG signal. An efficient feature extraction technique derives unique information about each movement hidden in the raw EMG signal [22,23]. To improve the EMG pattern recognition performance and ensure more degree of freedom, large numbers of time-domain, frequency-domain, and time-frequency-domain EMG features have been reported [24,25]. Popular time-domain features are found in many studies; these include the mean absolute value (MAV), waveform length (WL), number of zero crossings (ZC), and slope sign changes (SSC), which are mentioned in [26]; the variance (VAR), complexity (COM) and mobility (MOB), which are reported in [27]; and the Wilson amplitude (WAMP), log detector (LOG) and autoregressive coefficients (ARs), which are described in [28]. In addition, some other time-domain features are the myopulse percentage rate (MYOP) [29], skewness (SKW) [30], difference absolute mean value (DAMV) [31], difference absolute standard deviation value (DASDV) [31], root mean square (RMS) [24] and maximum fractal length (MFL) [32]. Thereafter, the most commonly used frequency-domain features are the mean frequency (MNF) and mean power (MNP) proposed in [22,33] and the median frequency (MDF), frequency ratio (FR), spectral moment

EMG signal reflects the activity of a muscle corresponding to a movement [5,6]. Electromyography is a technique that senses the bioelectrical potential, also known as the EMG signal, from a target muscle or group of muscles with the help of a surface electrode or needle electrode when these muscles are neurologically activated [7-10]. Generally, surface EMG is used for myoelectric

(SM), total power (TTP) and variance of the central frequency (VCF) given in [34]. Moreover, short-time Fourier transforms or wavelet transforms extract features both in time and frequency domains. However, in the literature, time-domain features are used frequently than frequency- or time-frequency-domain features since these features do not require any transformations and hence large computing resources, such as processing power, memory, etc. [25,29]. Consequently, this research is carried out to determine a further contribution to the time-domain feature set.

Khushaba *et al.* [35] proposed six time-domain features and spectrum correlations between each pair of channels for arm position-invariant EMG pattern recognition using a seven-channel EMG signal. Furthermore, Al-Timemy *et al.* [36] extended the work of Khushaba *et al.* [35] and introduced six modified time-domain features to improve EMG pattern recognition performance against muscle force variations, where eight channels were employed to collect EMG signals. Thereafter, Khushaba *et al.* [37] modified and upgraded their pilot work and proposed seven time-domain features that were validated over five EMG datasets. In these EMG datasets, the number of channels varied from eight to one hundred twenty-eight. Recently, Asogbon *et al.* [38] described five time-domain features to resolve the effects of both limb position and muscle force variation simultaneously, where they included eight EMG signal channels. The major limitation of a large number of existing features is that these features are unable to address all the requirements for a given application since a specific feature is effective for a specific type of application with a specific arrangement [26,28,29]. Again, all the authors mentioned above used multichannel EMG signals with a minimum of seven channels to validate their proposed features. However, multichannel myoelectric pattern recognition increases the computational cost and device cost [39]. In addition, some of the proposed features are applicable only for multichannel EMG systems [40-42]. Thus, the existing features validated for multichannel systems may not be effective for few numbers of channels. Again, a myoelectric pattern recognition system using the fewest possible channels lacks spatial information, which in turn decreases the separation margins among the movements [43-45]. Usually, this problem occurs among movements possessing relatively weaker signals than the others due to the overlapping of activated muscles, narrow muscles, muscle activations with low contraction forces, etc. [32,46].

In this context, to minimize these limitations, we propose two time-domain features, namely, the LMAV and NSV, which should improve the separation margins among the movements when the number of channels used is two. These proposed features are based on the nonlinear scaling, log, and cubic root of the signal amplitude. The LMAV produces relatively higher discrimination among weak signals than among strong signals. In addition, the NSV measures the nonlinear deviation of each sample value from its linear mean absolute value, focusing on the instantaneous amplitude of a weak signal. These proposed features maximize the margins among the least separable movements. Consequently, these proposed features improve the EMG pattern recognition performance of a model in terms of accuracy, sensitivity, specificity, precision and F1 score when these two features are grouped with the existing four feature extraction methods considered in this study. This performance improvement means that proposed, LMAV and NSV, add some new information to the existing feature extraction methods which interns contribute to improving EMG pattern recognition performance. Moreover, the LMAV and NSV show strength over variable window size, variable SNR, movement-wise performance enhancement and datasets. In addition, a combined feature extraction method is also proposed in this study, where the LMAV and NSV are grouped with existing 11 time-domain features, including the WL, WAMP, SSC, ZC, MOB, COM, SKW, and four autoregressive coefficients. The proposed feature extraction method achieves the highest EMG pattern recognition performance in terms of all performance evaluating parameters.

In terms of EMG pattern recognition, many classifiers have been used in recent studies. These are convolutional neural networks (CNNs) [47,48], linear discriminant analysis (LDA) [49], artificial neural networks (ANNs) [50], fuzzy methods [51], support vector machines (SVMs) [44,49], and k-nearest neighbours (KNNs) [52]. Among these methods, the CNN provides very strong EMG recognition performance but is impossible to implement in cheap hardware for real-time operation [53]. Therefore, to minimize the hardware cost and obtain an acceptable level of performance, we use LDA, an SVM, and the KNN algorithm as classifiers, all of which are widely used for these types of applications [37,49,52]. Furthermore, the resulting EMG pattern recognition performance is investigated over two datasets (newly collected and standard datasets with the same arrangements) to validate our results.

The rest of the sections are structured as follows. Section 2 describes the EMG datasets, proposed features, scatter plots and EMG pattern recognition method. Section 3 presents the experimental results, where the resulting performances are evaluated and compared with those of other considered feature extraction methods. Section 4 investigates the reasons for the obtained performance enhancement, and Section 5 concludes with the overall experimental results.

## 2. Methodology

### 2.1. EMG data collection

In this study, we employ two EMG datasets, where datasets 1 and 2 are collected using our EMG signal acquisition system and a public dataset from online, respectively.

### 2.1.1. Acquisition of dataset 1

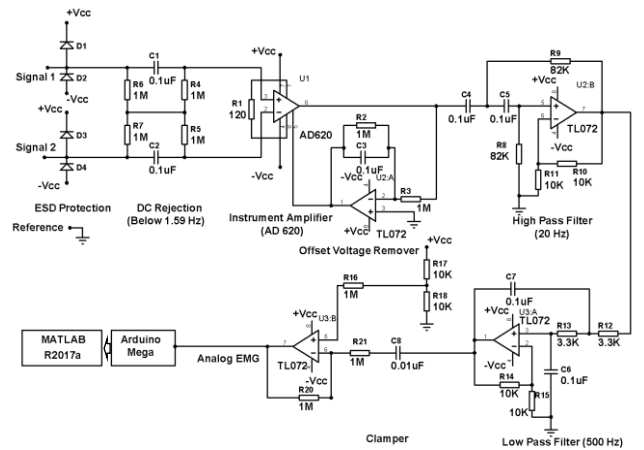

(a)

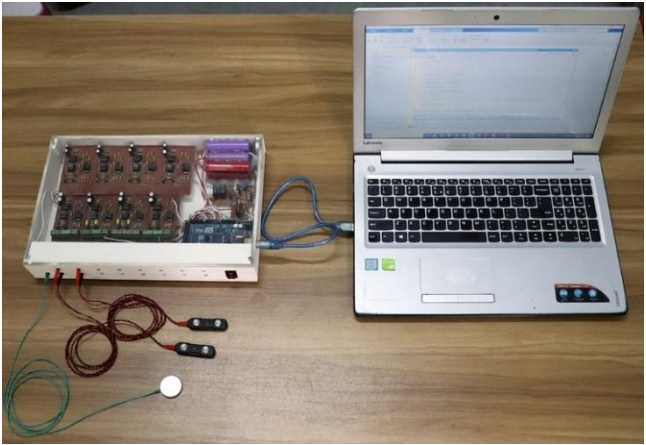

**Figure 1.** The EMG data acquisition system where (a) and (b) present the schematic circuit diagram and an EMG signal acquisition system, respectively.

For the acquisition of this EMG dataset, we employ an EMG signal acquisition system. Figure 1a shows the schematic circuit diagram of a single-channel bipolar EMG signal acquisition system. It consists of several functional blocks, including electrostatic discharge (ESD) protection, DC rejection, an instrument amplifier, a high-pass and low-pass filter, and a clamper. In this system, the ESD unit provides a low-resistance path for a high electrostatic charge on the human body and protects sophisticated devices [54]. In addition to electrostatic charge, the raw EMG signal also possesses DC half-cell potential produced on the electrode-skin interface and its amplitude is enough to saturate a high gain instrument amplifier. So, the DC offset voltage is removed by passing the signal through a DC rejection circuit (also known as a balanced ac coupling network). It is mainly a differential high-pass filter whose cut off frequency lies near to DC frequency. The advantage of this filter is that it offers a bias path without any ground connection resulting in a high common-mode rejection ratio (CMRR) of the instrument amplifier [55]. Thereafter, we use an instrument amplifier integrated circuit (AD620) for the differential amplification of the raw EMG signal [56]. In addition, there exists an offset voltage with regard to the instrument amplifier itself. To eliminate the offset voltage of the instrument amplifier, we employ a unity-gain inverting amplifier, and the output of this amplifier is used to ground the instrument amplifier [57]. During muscle contraction, the electrode shifts slightly which generate a noise also known as movement artefacts whose cut off frequency lies between 0 Hz to 20 Hz [58]. So, a second-order high-pass filter of 20 Hz is employed to remove it. Finally, we employ a second-order low-pass filter of 500 Hz to eliminate high-frequency noise [58]. Then, we use a positive clamper circuit to shift the DC level to 2.5 V. Finally, we employ an Arduino Mega for digitalizing the EMG signal at a resolution of 10 bits with a sampling frequency of 2000 Hz.

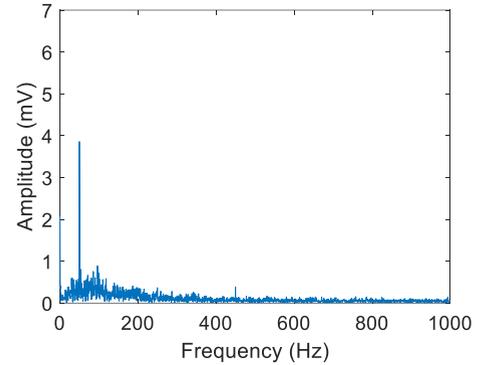

(a)

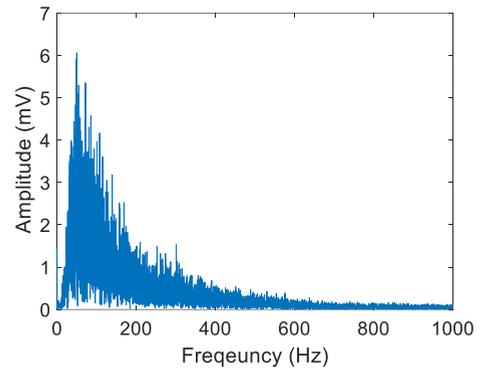

(b)

**Figure 2.** The frequency spectrum of EMG signal acquisition system where (a) noise and (b) EMG signal.

Figure 1b shows the implementation of the multichannel bipolar EMG signal acquisition system, where the circuit shown in figure 1a is repeated for all channels. In this system, we use an MFI bar electrode made in the USA as a surface electrode. In this device, the average EMG signal strength (RMS) during no movement condition is considered as noise and the value found is 11.9 mV where the system gain is 1652 (gain of instrument amplifier × gain of high pass filter × gain of low pass filter = 413×2×2). Figure 2a shows the frequency spectrum of noise where the dominant noise comes from power line artefacts and its harmonics, however, these can be minimized by employing a digital notch filter [59]. In addition to the power line artefact, figure 2a also indicates that the noise includes additive white Gaussian noise (AWGN) and some low amplitude EMG signal. Again, figure 2b shows the frequency spectrum of a movement (mixed with noise) which is very high in amplitude relative to the noise but it varies with the movements (figure 6) and muscle force levels [36]. In this dataset, the average SNR for medium force level is calculated using equation (1). First, noise (RMS) is eliminated from the RMS value of the raw EMG signal and the average SNR value for each subject is calculated in dB. In this study, the obtained SNR values vary across all subjects and lie between 5 dB to 23 dB.

$$SNR = 20\log_{10}(\frac{\sqrt{Raw\ EMG\ Signal_{RMS}^2 - Noise_{RMS}^2}}{Noise_{RMS}}) \quad (1)$$

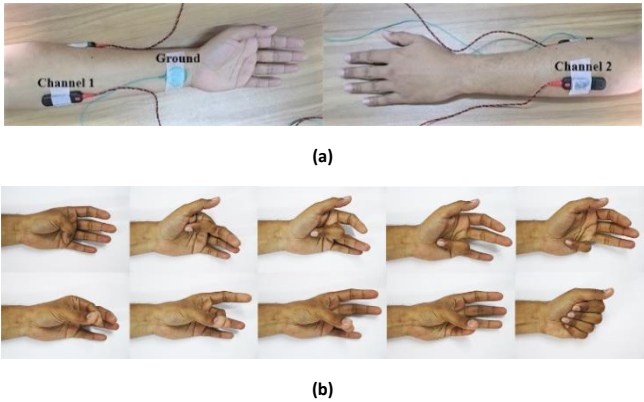

**Figure 3.** The EMG signal acquisition settings where (a) and (b) present the electrode placement and the finger movements, respectively.

For this dataset, we collected a two-channel EMG signal: channel 1 collects the EMG signal from the flexor digitorum superficialis, flexor digitorum profundus and remote flexor pollicis longus, and channel 2 collects the EMG signal from the extensor digitorum communis, extensor digiti minimi and remote extensor pollicis longus. In addition, the ground electrode was placed on the wrist, as shown in figure 3a. To ensure proper contact between the electrodes and skin, the electrodes were attached to the skin through an adhesive conductive gel. However, ten intact-limbed subjects aged between 25 to 55 years were engaged to perform five individual finger movements, thumb (T), index (I), middle (M), ring (R), little (L) and five combined finger movements, i.e., thumb-index (TI), thumb-middle (TM), thumb-ring (TR), thumb-little (TL) and hand closing (HC) movements, as shown in figure 3b. During this signal acquisition phase, we informed all the participants about the objective of the research, and they provided us with their written consent in this regard. Ethical approval was provided by the Faculty of Engineering, University of Rajshahi, Bangladesh to perform this study. During data collection, the subjects sat on a handled chair to place their hands freely. During the recording process, each of the finger movements was repeated six times with a duration of five seconds. In addition, the subjects were relaxed for 5 to 10 seconds between successive movements.

### 2.1.2. Description of dataset 2

In this study, we also collect the same dataset as dataset 1 to validate the result obtained from the Khushaba website [59]; here, Delsys DE 2.x series EMG sensors from Bagnoli Desktop EMG Systems are used for data acquisition. In this dataset, there are eight subjects, including six males and two females aged 20 to 35 years. Each subject performed five individual movements (T, I, M, R and L) and five combined finger movements (TI, TM, TR, TL and HC) as shown in figure 3b providing six trials for each movement where each trial is five seconds long in duration. The EMG signal is sampled at 4000 Hz and digitalized with a 12-bit resolution using a National Instruments BNC-2090. In this dataset, the signal is in raw condition with the significant frequency spectrum of 20 Hz to 500 Hz (Appendix A).

### 2.2. Feature extraction

An EMG signal is composed of hidden unique information for each movement. Feature extraction methods are employed to obtain as few features as possible, to obtain the most effective feature(s) or to derive a new feature(s) for a particular application. The performance of EMG pattern recognition strongly depends on proper feature selection rather than the classification algorithms used [29,34].

### 2.2.1. The proposed time-domain features

In our studies, we propose two time-domain features as described below:

The first proposal is the log of the mean absolute value (LMAV), which is mainly a nonlinear scaling of the mean absolute value. We find that the LMAV for a given window highly discriminates or focuses on a low amplitude EMG signal; for this reason, it is expected that it can provide better performances than currently used features. The LMAV is expressed mathematically as:

$$LMAV = Log_e(\sqrt{\frac{1}{N}\sum_{i=1}^{N}|x_i|}) \quad (2)$$

where $N$ represents the size of the window and $x_i$ denotes the $i^{th}$ sample within the corresponding window.

The second proposal is the nonlinear scaled value (NSV). This NSV is based on nonlinear scaling operations. The NSV measures the nonlinear deviation of each sample value from its linear mean absolute value to focus on the instantaneous amplitude of a weak EMG signal. It also emphasizes discriminating low amplitude EMG signals rather than high amplitude signals.

$$NSV = Log_e(\sqrt{\frac{1}{N}\sum_{i=1}^{N}(\overline{|x|} - |x_i|^{1/3})^2}) \quad (3)$$

Where $\overline{|x|}$ represents the mean absolute value for a window of size $N$.

### 2.2.2. The feature extraction methods

Time-domain features are widely used because they do not require any mathematical transformations or modifications; as a result, they ensure low time consumption in pattern recognition tasks [29]. In this study, we use five popular time-domain feature extraction methods, where each method includes several features. These include:

Huang et al. [60] used seven features (FS1) and six autoregressive coefficients along with the RMS value.

Du et al. [61] used six time-domain features (FS2) that include the integration of the EMG (IEMG), WL, WAMP, ZC, SSC, and VAR.

Time-dependent power spectrum descriptors (FS3) [36] introduce six time-domain features: the root squared zero-order, second-order and fourth-order moments; an irregularity factor; sparseness; and the waveform length ratio. In this study, we use six features directly.

The temporal-spatial descriptors (FS4) [37] describe seven time-domain features: the root squared zero-order, second-order and fourth-order moments; an irregularity factor; sparseness; the coefficient of variation; and the Teager-Kaiser energy operator. In this research, we employ seven features only.

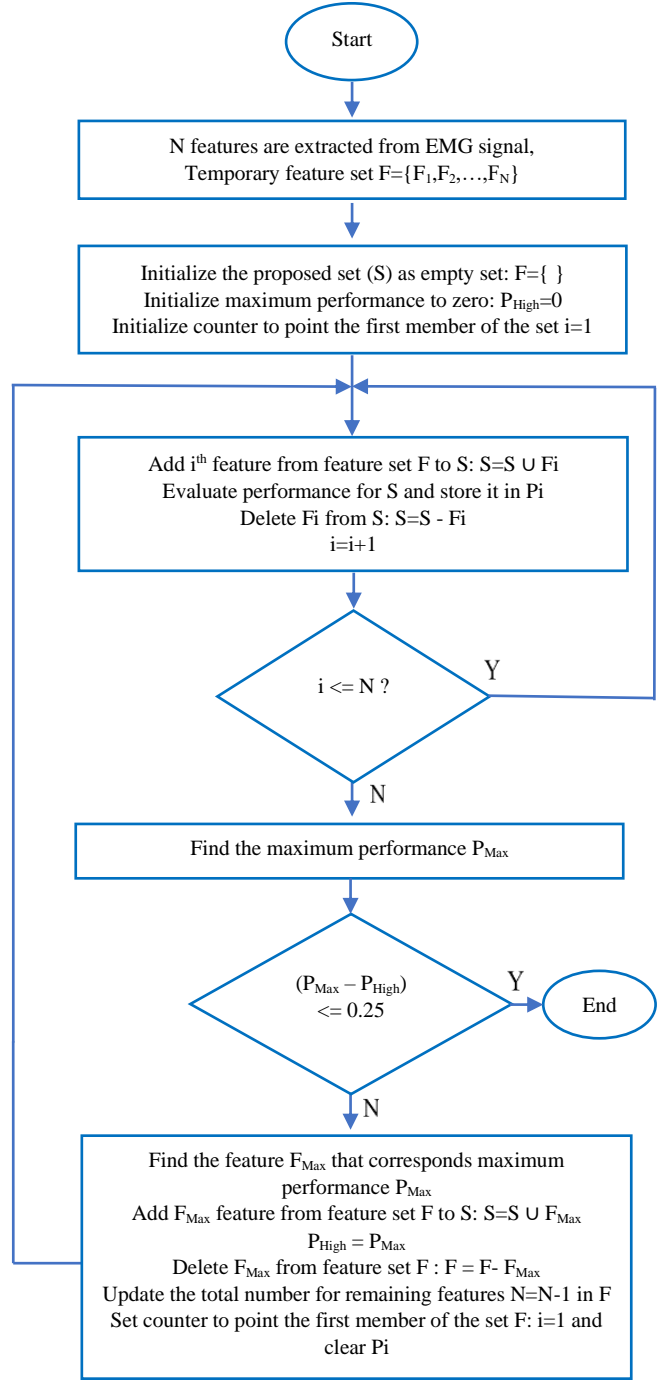

**Figure 4.** The forward feature selection algorithm.

In this study, we have proposed a combined feature extraction method which is selected from the proposed LMAV and NSV, and the existing time-domain features [26-32]. For the selection of these features, we use the forward feature selection algorithm [62] shown in figure 4, which is evaluated across 32 time-domain features. In this feature selection, first, we select the highest performing feature among all the features. Then, we group the best performing feature with each of the remaining features one by one and find the group that yields the strongest performance. However, we consider a new feature from the highest performing group only when it satisfies the condition of a minimum performance enhancement of 0.25. Thus, the algorithm selects 13 features which are denoted as the proposed feature extraction method, including the proposed LMAV and NSV along with existing 11 time-domain features i.e., WL, WAMP, SSC, ZC, MOB, COM, and SKW, and four autoregressive coefficients. Here, the WAMP gives the signal energy, the WL provides the collective length of the EMG waveform, the SSC and ZC describe indirect frequency information, the MOB represents the mean frequency or the proportion of the standard deviation of the power spectrum, the COM measures the change in frequency, the SKW is the degree of asymmetry of the spreading of a random variable around the variable mean and the AR coefficients are based on the linear predictive model.

### 2.3. Scatter plot

A scatter plot is normally a presentation of two variables calculated from a dataset using Cartesian coordinates, and this plot is used to visually observe the clustering performance of an algorithm and the degree of overlap among classes. The selection of two variables depends on our choice: unique variables from two channels, two separate variables from a single channel, or the first two reduced features obtained from the dimension reduction technique [29]. In our scatter plot, we use two reduced features from uncorrelated linear discriminant analysis (ULDA). Here, subject 1 data from dataset 1 is used, and each data point on the scatter plot denotes the first two reduced features from ULDA. In this case, the size of the window considered is 250 ms. Therefore, the ten different movements are presented by three hundred data points (data points × trials × movements = 5×6×10) on a single scatter plot. Moreover, the reduced feature values are normalized by using equation (4) to obtain a better presentation [29].

$$Normfeat_i = \frac{feat_i - \min_i}{\max_i - \min_i} \qquad (4)$$

where $max_i$ and $min_i$ are the maximum and minimum values of the $i^{th}$ feature, respectively.

### 2.4. RES index

For evaluating the clustering performances of features, the statistical parameter RES (ratio of the Euclidean distance to the standard deviation) index is used. A higher RES index indicates higher separation among the classes and vice versa. The benefit of using the RES index is that it is independent of the classifiers used. The RES index can be evaluated as follows [63]:

$$RES\ Index = \frac{\overline{ED}}{\overline{\sigma}} \qquad (5)$$

where $\overline{ED}$ is the Euclidean distance between movements $p$ and $q$. It is defined mathematically as

$$\overline{ED} = \frac{2}{K(K-1)} \sum_{p=1}^{K-1} \sum_{q=p+1}^{K} \sqrt{(m_{1p} - m_{1q})^2 + (m_{2p} - m_{2q})^2} \qquad (6)$$

where $m$ is the mean value of a feature and $k$ denotes the total number of movements. The dispersion of clusters $p$ and $q$ is given by

$$\overline{\sigma} = \frac{1}{IK} \sum_{i=1}^{I} \sum_{k=1}^{K} s_{ik} \qquad (7)$$

where $l$ is the length of the feature vector.

### 2.5. EMG pattern recognition method

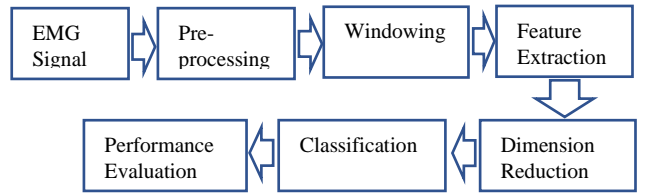

**Figure 5.** Block diagram of the myoelectric pattern recognition system.

Figure 5 shows the block diagram of the myoelectric pattern recognition system, where we employ MATLAB R2017a software (Mathworks, USA) for the EMG pattern recognition of ten-finger movements. After getting the digital EMG signal through the process in figure 1a or from dataset 2, we pass the EMG signal through a digital preprocessing block using MATLAB R2017a environment where a digital bandpass filter (20 to 500 Hz) and a digital notch filter (50 Hz) are used to reduce movement artefacts, high-frequency noise [58], and power line artefacts [64]. In general, two types of windowing, i.e., overlapped and disjoint windowing, are used [65]; however, between these two, the overlapped windowing scheme offers better pattern recognition performance, but its computational cost is higher [66]. As a result, to obtain a lower computational cost with simplicity [59], we use a 250 ms disjoint windowing scheme [67] that produces 20 segments for each five-second long dataset. Hence, features are extracted using feature extraction methods (i.e., FS1, FS2, FS3, and FS4) that create a high-dimensional feature space, as mentioned in Section 2.2.2. The feature dimensionality is reduced (total classes – 1 = 10 - 1 = 9) using ULDA [68]. Now, the 9-dimensional reduced feature vectors for each feature extraction method are classified using three popular classifiers: LDA with quadratic function [69,70], SVM with gaussian radian basis kernel function (sigmavalue=1) [71], and KNN with cityblock distance (neighbours=3) [35]. In this performance evaluation, five trials containing 1000 samples (trials × movements × samples per trial = 5×10×20) are used as training data, and the remaining trial containing 200 samples (trials × movements × samples per trial = 1×10×20) is used as testing data. The process is repeated six times so that each of the trials is employed as testing data like 6-fold cross-validation where trial-wise performance evaluation is performed according to [36,37]. In addition to generating a large number of training samples, the EMG pattern recognition performances during the training and testing periods are also compared, and it is found that the differences in their performances are negligible, which implies that the data are not overfitted. However, EMG pattern recognition performance is measured by accuracy, sensitivity, specificity, precision, and the F1 score [72,73]. Accuracy, sensitivity, specificity, and precision describe the ability of a model to distinguish true positive and true negative movements; these metrics represent the number of positive movements correctly identified as positive, the number of negative movements correctly identified as negative, and the number of true positive movements over the positive predicted movements, respectively. Additionally, the F1 score combines both sensitivity and precision to find the true positive movements more precisely. These performance evaluation parameters can be defined as follows:

$$Accuracy = \frac{TP+TN}{TP+TN+FP+FN} \qquad (8)$$

$$Sensitivity = \frac{TP}{TP+FN} \qquad (9)$$

$$Specificity = \frac{TN}{TN+FP} \qquad (10)$$

$$Precision = \frac{TP}{TP+FP} \qquad (11)$$

$$F1\ Score = \frac{2 \times Precision \times Sensitivity}{Precision + Sensitivity} \qquad (12)$$

where $TP$, $TN$, $FP$, and $FN$ denote true positive movements, true negative movements, false positive movements, and false negative movements, respectively.

### 2.6. Statistical analysis

To find the significant differences between any pairs of feature extraction methods mentioned in Section 2.2.2, a Bonferroni corrected analysis of variance (ANOVA) test is utilized with a significance level of 0.05. The obtained *p*-values below 0.05 imply that the performances are significantly different. In this study, the EMG pattern recognition performances on both datasets are concatenated to construct an 18-dimensional vector (10 and 8 subjects in dataset 1 and dataset 2, respectively), and then a Bonferroni corrected ANOVA test is performed.

## 2. Results

### 3.1. Signal observation

The EMG signals in the time domain for the ten individual and combined finger movements collected from forearm muscles are presented in figure 6. In this figure, a time span of 250 ms is used for each finger movement. Here, no distinguishable features except amplitude are visually observed. Additionally, it is quite impossible to discriminate all movements successfully using either a single channel or a single feature. Therefore, in general, complex mathematical functions or transformations are used to enhance EMG pattern recognition performance for a minimal number of channels used.

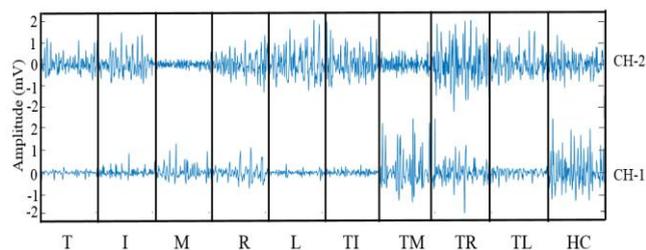

**Figure 6.** The EMG signal of dataset 2 in the time domain for ten-finger movements collected from two channels.

## 3.2. Scatter plot and RES index

The scatter plot for the different feature extraction methods mentioned in Section 2.2.2 is shown in figure 7. In this scatter plot, all the features of the respective feature extraction methods are extracted, and the obtained high-dimensional feature space is reduced by employing ULDA. Then, ULDA features 1 and 2 are plotted in the horizontal and vertical directions, respectively.

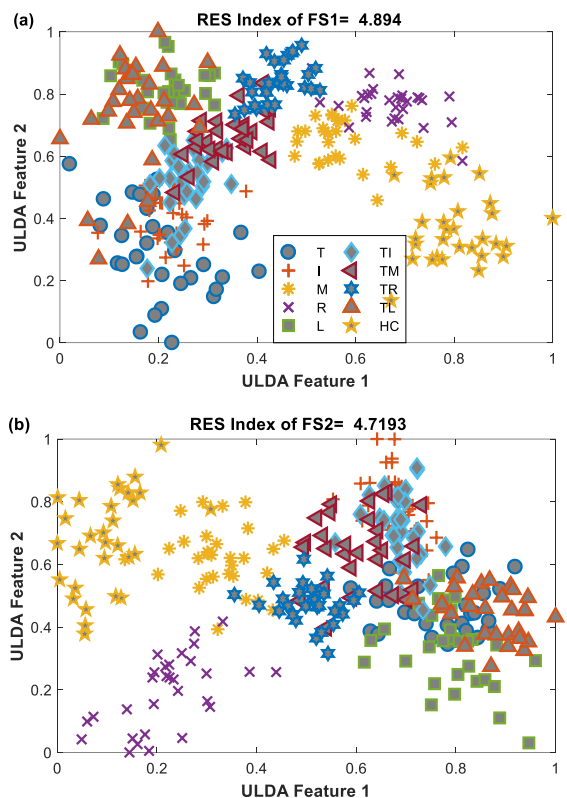

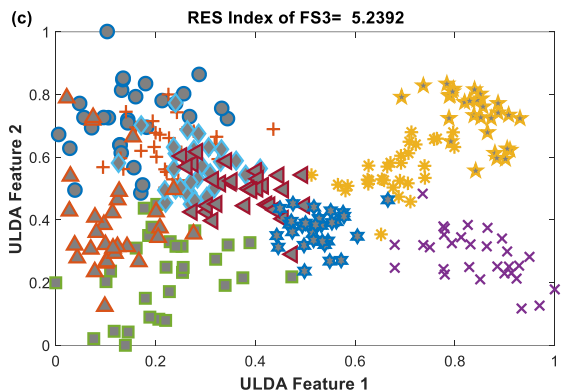

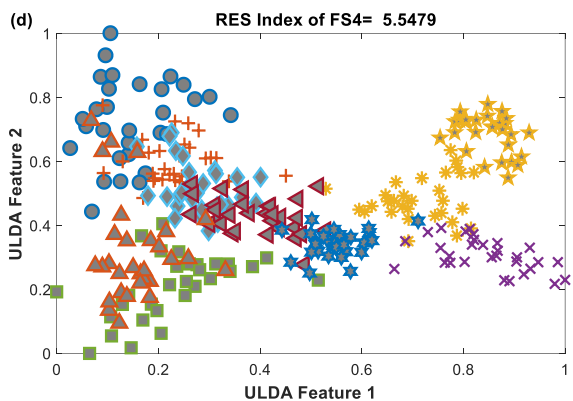

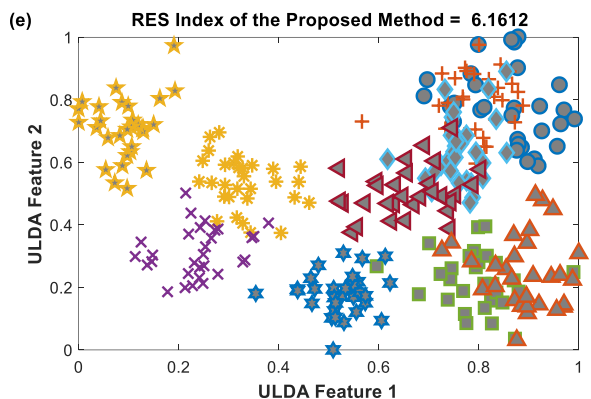

**Figure 7.** The scatter plot and RES index of different feature extraction methods for subject 1 of dataset 1 where (a) FS1, (b) FS2, (c) FS3, (d) FS4, and (e) the proposed method.

Figure 7 shows that the proposed feature extraction method provides better clustering performance than the existing feature extraction methods considered. ULDA features 1 and 2 are also used to calculate the RES index shown with the title of the corresponding scatter plot. The obtained results also indicate that the proposed feature extraction method provides the highest RES index compared to four existing feature extraction methods. Therefore, it is expected that the proposed feature extraction method could provide the best EMG pattern recognition performance.

### 3.3. EMG pattern recognition performance

To find the strength of the proposed feature extraction method in EMG pattern recognition performance, we compare the performances of the proposed feature extraction method with four existing feature extraction methods (FS1, FS2, FS3 and FS4). The comparison among the feature extraction methods in terms of accuracy, sensitivity, specificity, precision and F1 score is shown in Table 1.

**Table 1:** The EMG pattern recognition performances of different feature extraction methods.

| | Parameter | Classifier | FS1 | FS2 | FS3 | FS4 | Proposed |
|---|---|---|---|---|---|---|---|
| Dataset 1 | Accuracy | LDA | 96.79±1.00 | 96.27±0.88 | 96.21±0.92 | 96.03±1.05 | 97.79±0.52 |
| | | SVM | 95.73±0.93 | 95.14±1.06 | 95.53±0.93 | 95.66±1.02 | 97.52±0.56 |
| | | KNN | 94.94±1.04 | 94.39±1.12 | 94.95±0.98 | 95.25±1.15 | 97.23±0.63 |
| | Sensitivity | LDA | 83.96±4.98 | 81.36±4.38 | 81.03±4.60 | 80.15±5.25 | 88.97±2.58 |
| | | SVM | 78.65±4.67 | 75.70±5.30 | 77.63±4.64 | 78.30±5.08 | 87.62±2.80 |
| | | KNN | 74.71±5.19 | 71.94±5.60 | 74.74±4.90 | 76.24±5.77 | 86.16±3.13 |
| | Specificity | LDA | 98.22±0.55 | 97.93±0.49 | 97.89±0.51 | 97.79±0.58 | 98.77±0.29 |
| | | SVM | 97.63±0.52 | 97.30±0.59 | 97.51±0.52 | 97.59±0.56 | 98.46±0.35 |
| | | KNN | 97.19±0.58 | 96.88±0.62 | 97.19±0.54 | 97.36±0.64 | 98.62±0.31 |
| | Precision | LDA | 85.24±4.72 | 82.41±4.28 | 82.55±4.49 | 81.41±4.32 | 89.95±2.53 |
| | | SVM | 79.42±5.00 | 77.11±5.13 | 78.83±4.81 | 79.61±5.35 | 88.47±2.77 |
| | | KNN | 75.96±5.19 | 73.59±5.40 | 76.27±4.69 | 77.74±5.58 | 87.03±3.11 |
| | F1 Score | LDA | 83.49±5.13 | 80.56±4.57 | 80.58±4.69 | 79.41±5.59 | 88.55±2.72 |
| | | SVM | 78.08±4.84 | 75.06±5.41 | 76.97±4.83 | 77.71±5.32 | 87.26±2.88 |
| | | KNN | 74.22±5.26 | 71.38±5.64 | 74.26±4.94 | 75.74±5.91 | 85.77±3.24 |
| Dataset 2 | Accuracy | LDA | 97.18±0.94 | 96.73±0.93 | 96.69±1.07 | 95.78±1.08 | 98.36±0.45 |
| | | SVM | 96.58±0.97 | 96.00±0.85 | 95.88±1.00 | 95.32±1.29 | 98.07±0.59 |
| | | KNN | 96.00±1.08 | 95.54±0.96 | 95.47±1.24 | 94.85±1.41 | 97.97±0.66 |
| | Sensitivity | LDA | 85.88±4.72 | 83.66±4.66 | 83.43±5.34 | 78.88±5.40 | 91.78±2.25 |
| | | SVM | 82.88±4.86 | 80.01±4.24 | 79.42±4.99 | 76.60±6.44 | 90.34±2.96 |
| | | KNN | 80.06±5.38 | 77.69±4.80 | 77.34±6.20 | 74.25±7.04 | 89.83±3.28 |
| | Specificity | LDA | 98.43±0.52 | 98.18±0.52 | 98.16±0.59 | 97.65±0.60 | 99.09±0.25 |
| | | SVM | 98.10±0.54 | 97.78±0.47 | 97.71±0.55 | 97.40±0.72 | 98.93±0.33 |
| | | KNN | 97.78±0.60 | 97.52±0.53 | 97.48±0.69 | 97.14±0.78 | 98.87±0.36 |
| | Precision | LDA | 87.02±4.69 | 85.26±4.31 | 84.79±4.86 | 80.05±4.90 | 92.65±1.99 |
| | | SVM | 83.92±4.92 | 81.51±4.05 | 80.75±5.01 | 78.12±6.58 | 91.20±2.68 |
| | | KNN | 81.62±5.41 | 79.11±4.71 | 78.68±6.11 | 75.36±7.28 | 90.70±2.95 |
| | F1 Score | LDA | 85.55±4.84 | 83.22±4.68 | 83.14±5.34 | 78.37±5.50 | 91.59±2.29 |
| | | SVM | 82.63±5.01 | 79.78±4.34 | 79.00±5.21 | 76.17±6.70 | 90.19±3.02 |
| | | KNN | 79.87±5.58 | 77.30±5.12 | 77.00±6.40 | 73.71±7.33 | 89.70±3.39 |

The table indicates that the proposed feature extraction method achieves the highest EMG pattern recognition performance in terms of all performance evaluating parameters. In this study, the FS1 achieves the second-best performance. However, if we compare this FS1 and the proposed feature extraction method, we find that on dataset 1 with LDA classifier, the proposed feature extraction method improves accuracy, sensitivity, specificity, precision, and F1 score by 1.00%, 5.01%, 0.55%, 4.71%, and 5.06%, respectively; again, on dataset 2 with LDA classifier, these improvements in accuracy, sensitivity, specificity, precision, and F1 score are 1.18%, 5.90%, 0.66%, 5.63%, and 6.04%, respectively. The higher F1 score, as found, indicates that the true positive for movements recognition rate is higher, which is generally expected. In addition, the *p*-value between the proposed method and FS1 is less than 0.001 considering all cases (Appendix B, Table B1). So, the lowest *p*-values indicate that the proposed method significantly improves EMG pattern recognition performance. Also, the comparison is shown graphically in figure 8 where only the F1 score is used for simple presentation.

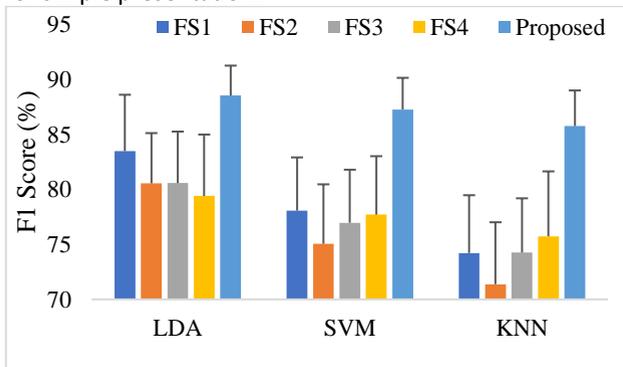

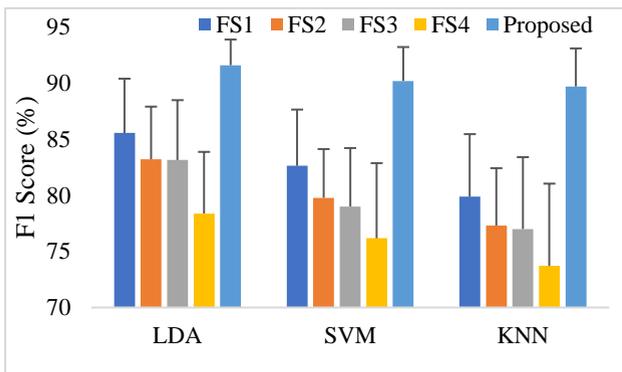

**Figure 8.** The F1 score of different feature extraction methods where (a) Dataset 1 and (b) Dataset 2

### 3.4. Performance enhancement of existing feature extraction methods with the LMAV and NSV

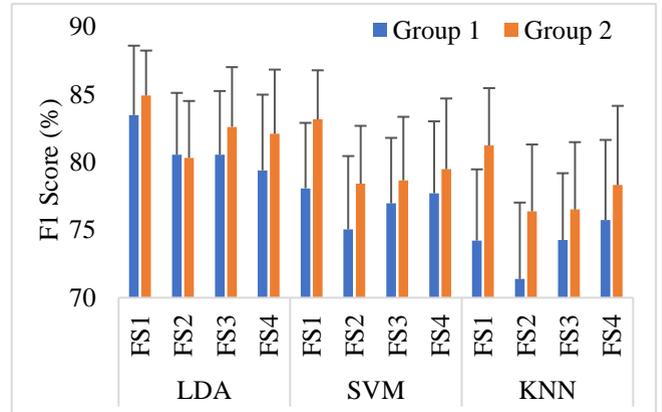

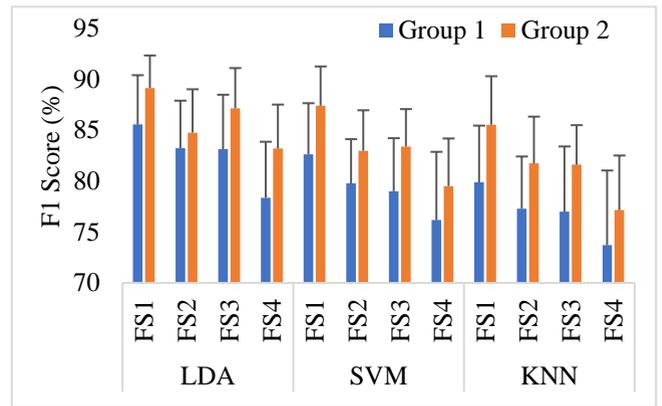

**Figure 9.** The F1 score enhancement of existing feature extraction methods with the LMAV and NSV where (a) Dataset 1 and (b) Dataset 2.

To demonstrate the effects of the proposed features, the LMAV and NSV, on EMG pattern recognition performances, the considered feature extraction methods are arranged into two groups: group 1 contains all the existing feature extraction methods (FS1, FS2, FS3, and FS4) mentioned in Section 2.2.2, and group 2 contains each of the existing feature extraction methods along with the LMAV and NSV. In this performance evaluation, we employ LDA, SVM and KNN classifiers and their experimental results are shown in Table 2a, Table 2b and Table 2c, respectively. Again, to show this performance enhancement with simplicity, we consider the F1 score for both datasets, as shown in figure 9. The tables imply that the LMAV and NSV enhance the EMG pattern recognition performances of the existing four feature extraction methods with three classifiers except for a negligible degradation in the performance of FS2 with the LDA classifier on dataset 1. The performance enhancement induced by using the LMAV and NSV for each feature extraction method is also validated with a Bonferroni-corrected ANOVA. The highest *p*-value between group 1 and group 2 is 0.002 (Appendix B, Table B2) considering all cases except for FS2 with LDA. The obtained *p*-values indicate that the LMAV and NSV significantly enhance EMG pattern recognition performance.

**Table 2a:** The EMG pattern recognition performance enhancement of existing feature extraction methods by the LMAV and NSV when LDA classifier is used.

|  | Parameter | Group | FS1 | FS2 | FS3 | FS4 |
|---|---|---|---|---|---|---|
| Dataset 1 | Accuracy | Group 1 | 96.79±1.00 | 96.27±0.88 | 96.21±0.92 | 96.03±1.05 |
|  |  | Group 2 | 97.10±0.62 | 96.23±0.81 | 96.62±0.86 | 96.54±0.91 |
|  | Sensitivity | Group 1 | 83.96±4.98 | 81.36±4.38 | 81.03±4.60 | 80.15±5.25 |
|  |  | Group 2 | 85.48±3.09 | 81.16±4.05 | 83.10±4.28 | 82.68±4.55 |
|  | Specificity | Group 1 | 98.22±0.55 | 97.93±0.49 | 97.89±0.51 | 97.79±0.58 |
|  |  | Group 2 | 98.39±0.34 | 97.91±0.45 | 98.12±0.48 | 98.08±0.51 |
|  | Precision | Group 1 | 85.24±4.72 | 82.41±4.28 | 82.55±4.49 | 81.41±4.32 |
|  |  | Group 2 | 86.89±3.09 | 82.66±4.00 | 84.54±4.20 | 84.00±4.48 |
|  | F1 Score | Group 1 | 83.49±5.13 | 80.56±4.57 | 80.58±4.69 | 79.41±5.59 |
|  |  | Group 2 | 84.94±3.32 | 80.33±4.21 | 82.62±4.42 | 82.10±4.75 |
| Dataset 2 | Accuracy | Group 1 | 97.18±0.94 | 96.73±0.93 | 96.69±1.07 | 95.78±1.08 |
|  |  | Group 2 | 97.87±0.64 | 97.02±0.84 | 97.48±0.78 | 96.71±0.86 |
|  | Sensitivity | Group 1 | 85.88±4.72 | 83.66±4.66 | 83.43±5.34 | 78.88±5.40 |
|  |  | Group 2 | 89.34±3.19 | 85.08±4.20 | 87.39±3.91 | 83.53±4.28 |
|  | Specificity | Group 1 | 98.43±0.52 | 98.18±0.52 | 98.16±0.59 | 97.65±0.60 |
|  |  | Group 2 | 98.82±0.35 | 98.34±0.47 | 98.60±0.43 | 98.17±0.48 |
|  | Precision | Group 1 | 87.02±4.69 | 85.26±4.31 | 84.79±4.86 | 80.05±4.90 |
|  |  | Group 2 | 90.49±2.66 | 86.51±3.78 | 88.60±3.53 | 84.96±4.07 |
|  | F1 Score | Group 1 | 85.55±4.84 | 83.22±4.68 | 83.14±5.34 | 78.37±5.50 |
|  |  | Group 2 | 89.14±3.19 | 84.76±4.26 | 87.14±3.95 | 83.19±4.32 |

**Table 2b:** The EMG pattern recognition performance enhancement of existing feature extraction methods by the LMAV and NSV when SVM classifier is used.

|  | Parameter | Group Set | FS1 | FS2 | FS3 | FS4 |
|---|---|---|---|---|---|---|
| Dataset 1 | Accuracy | Group 1 | 95.73±0.93 | 95.14±1.06 | 95.53±0.93 | 95.66±1.02 |
|  |  | Group 2 | 96.74±0.69 | 95.81±0.84 | 95.84±0.90 | 96.02±1.00 |
|  | Sensitivity | Group 1 | 78.65±4.67 | 75.70±5.30 | 77.63±4.64 | 78.30±5.08 |
|  |  | Group 2 | 83.68±3.43 | 79.06±4.20 | 79.19±4.50 | 80.10±5.02 |
|  | Specificity | Group 1 | 97.63±0.52 | 97.30±0.59 | 97.51±0.52 | 97.59±0.56 |
|  |  | Group 2 | 97.97±0.46 | 97.44±0.54 | 97.44±0.55 | 97.64±0.65 |
|  | Precision | Group 1 | 79.42±5.00 | 77.11±5.13 | 78.83±4.81 | 79.61±5.35 |
|  |  | Group 2 | 84.59±3.55 | 80.28±3.91 | 80.21±4.71 | 81.20±5.11 |
|  | F1 Score | Group 1 | 78.08±4.84 | 75.06±5.41 | 76.97±4.83 | 77.71±5.32 |
|  |  | Group 2 | 83.18±3.62 | 78.43±4.27 | 78.67±4.69 | 79.49±5.22 |

|  | Parameter | Group | FS1 | FS2 | FS3 | FS4 |
|---|---|---|---|---|---|---|
| Dataset 2 | Accuracy | Group 1 | 96.58±0.97 | 96.00±0.85 | 95.88±1.00 | 95.32±1.29 |
|  |  | Group 2 | 97.51±0.76 | 96.63±0.78 | 96.73±0.71 | 95.95±0.92 |
|  | Sensitivity | Group 1 | 82.88±4.86 | 80.01±4.24 | 79.42±4.99 | 76.60±6.44 |
|  |  | Group 2 | 87.53±3.79 | 83.13±3.90 | 83.63±3.56 | 79.75±4.59 |
|  | Specificity | Group 1 | 98.10±0.54 | 97.78±0.47 | 97.71±0.55 | 97.40±0.72 |
|  |  | Group 2 | 98.61±0.42 | 98.13±0.43 | 98.18±0.40 | 97.75±0.51 |
|  | Precision | Group 1 | 83.92±4.92 | 81.51±4.05 | 80.75±5.01 | 78.12±6.58 |
|  |  | Group 2 | 88.56±3.49 | 84.38±3.51 | 84.85±3.38 | 81.33±4.38 |
|  | F1 Score | Group 1 | 82.63±5.01 | 79.78±4.34 | 79.00±5.21 | 76.17±6.70 |
|  |  | Group 2 | 87.40±3.85 | 82.98±3.98 | 83.39±3.68 | 79.51±4.67 |

**Table 2c:** The EMG pattern recognition performance enhancement of existing feature extraction methods by the LMAV and NSV when KNN classifier is used.

|  | Parameter | Group | FS1 | FS2 | FS3 | FS4 |
|---|---|---|---|---|---|---|
| Dataset 1 | Accuracy | Group 1 | 94.94±1.04 | 94.39±1.12 | 94.95±0.98 | 95.25±1.15 |
|  |  | Group 2 | 96.35±0.82 | 95.39±0.97 | 95.40±0.99 | 95.76±1.17 |
|  | Sensitivity | Group 1 | 74.71±5.19 | 71.94±5.60 | 74.74±4.90 | 76.24±5.77 |
|  |  | Group 2 | 81.76±4.11 | 76.97±4.84 | 76.98±4.97 | 78.80±5.86 |
|  | Specificity | Group 1 | 97.19±0.58 | 96.88±0.62 | 97.19±0.54 | 97.36±0.64 |
|  |  | Group 2 | 98.19±0.38 | 97.67±0.47 | 97.69±0.50 | 97.79±0.56 |
|  | Precision | Group 1 | 75.96±5.19 | 73.59±5.40 | 76.27±4.69 | 77.74±5.58 |
|  |  | Group 2 | 83.04±3.97 | 78.21±4.69 | 78.43±4.71 | 80.10±5.53 |
|  | F1 Score | Group 1 | 74.22±5.26 | 71.38±5.64 | 74.26±4.94 | 75.74±5.91 |
|  |  | Group 2 | 81.25±4.23 | 76.39±4.93 | 76.52±4.97 | 78.32±5.85 |
| Dataset 2 | Accuracy | Group 1 | 96.00±1.08 | 95.54±0.96 | 95.47±1.24 | 94.85±1.41 |
|  |  | Group 2 | 97.15±0.91 | 96.40±0.88 | 96.36±0.76 | 95.49±1.04 |
|  | Sensitivity | Group 1 | 80.06±5.38 | 77.69±4.80 | 77.34±6.20 | 74.25±7.04 |
|  |  | Group 2 | 85.74±4.54 | 82.00±4.38 | 81.80±3.79 | 77.47±5.19 |
|  | Specificity | Group 1 | 97.78±0.60 | 97.52±0.53 | 97.48±0.69 | 97.14±0.78 |
|  |  | Group 2 | 98.42±0.50 | 98.00±0.49 | 97.98±0.42 | 97.50±0.58 |
|  | Precision | Group 1 | 81.62±5.41 | 79.11±4.71 | 78.68±6.11 | 75.36±7.28 |
|  |  | Group 2 | 86.81±4.19 | 83.36±4.24 | 83.17±3.55 | 79.15±4.54 |
|  | F1 Score | Group 1 | 79.87±5.58 | 77.30±5.12 | 77.00±6.40 | 73.71±7.33 |
|  |  | Group 2 | 85.54±4.76 | 81.73±4.60 | 81.61±3.88 | 77.15±5.37 |

## 3.5. Movement-wise performance enhancement induced by using the LMAV and NSV

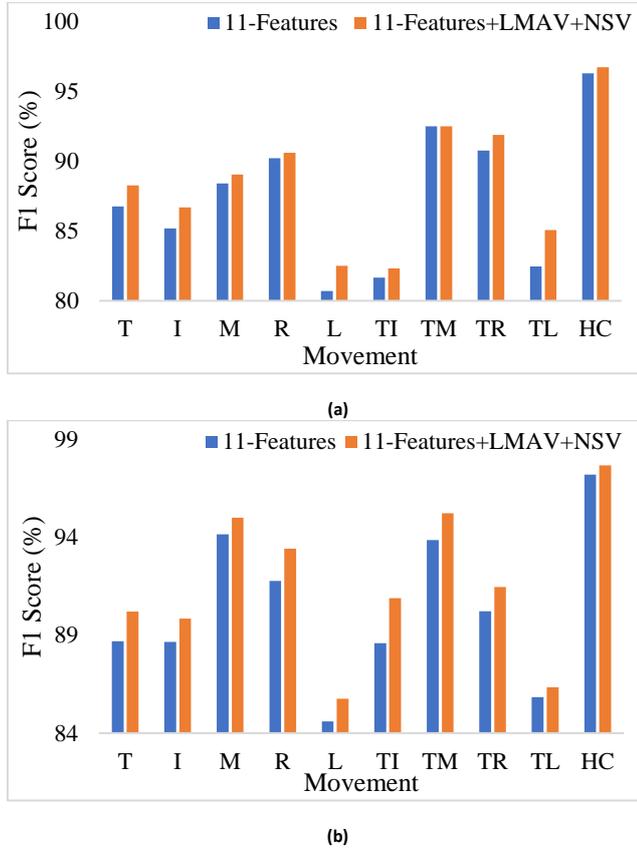

**Figure 10.** Movement-wise performance enhancement by using the LMAV and NSV using LDA classifier where (a) dataset 1 and (b) dataset 2.

To determine the movement-wise performance enhancement induced by using the LMAV and NSV, we consider the proposed feature extraction method which includes the proposed LMAV and NSV and existing 11 time-domain features as described in Section 2.2.2. Furthermore, the EMG pattern recognition performance (F1 score) is evaluated for existing 11 time-domain features and existing 11 time-domain features along with LMAV and NSV. In this performance evaluation, we employ an LDA classifier with a 250 ms window. Figure 10a and figure 10b show the performance for dataset 1 and dataset 2, respectively. The figures demonstrate that the LMAV and NSV improve the F1 score for all the cases except for TM and HC movement of dataset 1. Moreover, all hand movements achieve a noticeably better F1 score (up to 2.60% and 2.30% for dataset 1 and dataset 2, respectively) for the proposed feature extraction method, the LMAV and NSV, along with existing 11 time-domain features. Also, the obtained *p*-value between the overall performances of 11-features and 11-features along with the LMAV and NSV is less than 0.001 which indicates the significant improvement by the LMAV and NSV.

## 2.6. Impact of the LMAV and NSV on performance enhancement with a variable window size

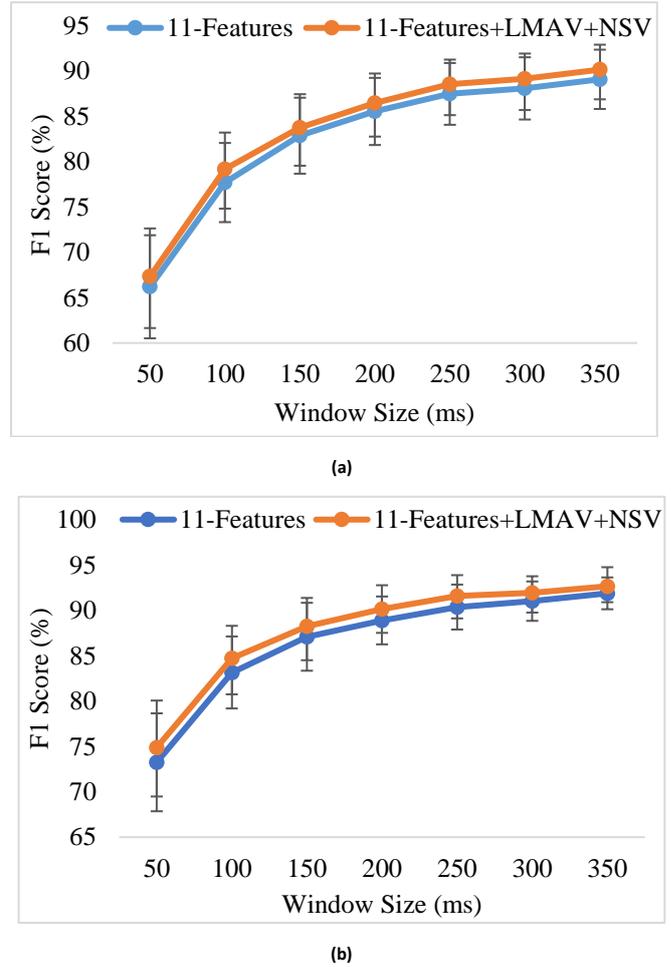

**Figure 11.** The impact of the LMAV and NSV on F1 Score with variable window size using LDA classifier where (a) dataset 1 and (b) dataset 2.

To investigate the impact of the LMAV and NSV on the performance enhancement with variable window size, we vary the window size from 50 ms to 350 ms with an interval of 50 ms. Then, we consider the proposed feature extraction method only where the F1 score is evaluated for existing 11 time-domain features and existing 11 time-domain features along with LMAV and NSV using the LDA classifier, as shown in figure 11. The figure indicates that the LMAV and NSV improve the F1 score for all window sizes across both datasets. In addition to the F1 score, the other performance evaluation parameters (accuracy, sensitivity, specificity, and precision) follow the trend of the F1 score. In this study, the SVM and KNN also provide a similar set of consistent result compared to those obtained under LDA. Moreover, it is also noted that the standard deviation decreases with increasing window size. A similar phenomenon is also observed for the other two classifiers. In addition, we evaluate *p*-values between the performances of 11-features and 11-features along with LMAV and NSV for each window size. The obtained *p*-values between the F1 scores of 11-features and 11-features along with LMAV and NSV at various window size are less than 0.001 (Appendix B, Table B3). The smallest *p*-values indicate that the LMAV and NSV significantly improve EMG pattern recognition performance for variable window size.

## 2.7. Impact of the LMAV and NSV on performance enhancement with a variable SNR

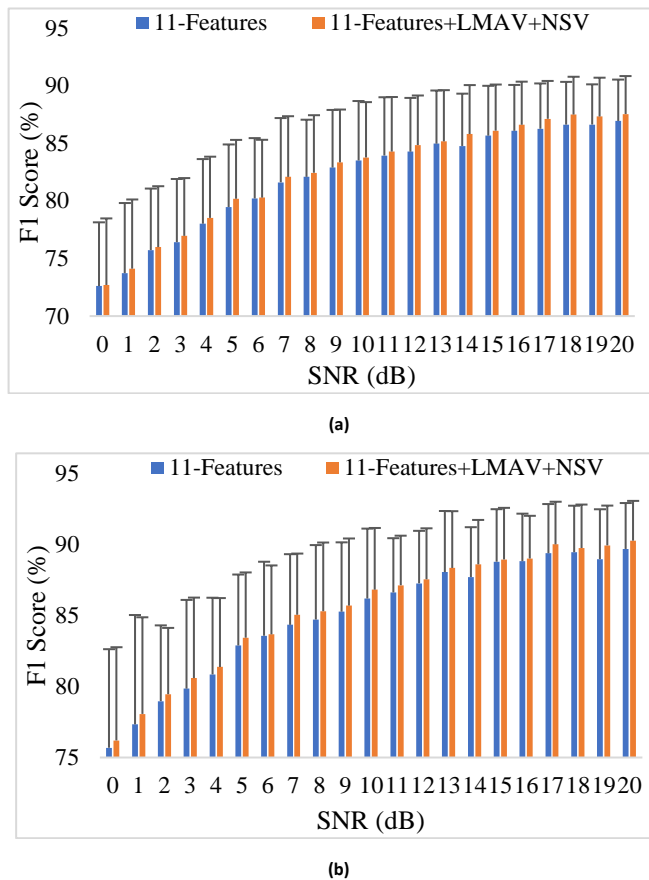

**Figure 12.** The impact of the LMAV and NSV on F1 with variable SNR Score using LDA classifier where (a) dataset 1 and (b) dataset 2.

To find the strength of the proposed LMAV and NSV with variable SNR, we consider the proposed feature extraction method where we mix AWGN artificially to the raw EMG signal which ranges from 0 dB to 20 dB with an interval of 1 dB [74,75]. In this noise mixing, we employ MATLAB R2017a function (*awgn*) to generate an AWGN of specific dB with respect to signal and to mix the noise with the signal. Then, the EMG pattern recognition performance (F1 score) is measured for the existing 11 time-domain features and the existing 11 time-domain features along with LMAV and NSV using an LDA classifier. The EMG pattern recognition performance with standard deviation is shown in figure 12. The following figures indicate that the EMG pattern recognition performance of time-domain features is affected by different level of AWGN; but the performance is almost stable above the SNR value of 16 dB. Another important point noted that the proposed LMAV and NSV contribute more or less to enhance the performance for all SNR values and it is valid for both datasets. Another important point noted that the proposed LMAV and NSV significantly contribute to enhancing the EMG pattern recognition performance from the SNR value of 17 dB to 20 dB since the *p*-values are less than 0.01 (Appendix B, Table B4).

## 4. Discussion

Bio-signals, such as those from electroencephalograms (EEGs), EMGs, electrocardiograms (ECGs) and photoplethysmograms (PPGs), have been widely investigated in the diagnosis of diseases and the real-time monitoring of patients [76-79]. Among these bio-signals, EMG signals are widely studied as a control signal for prosthetic hands [9]. However, there is an industrial demand for providing amputees with a low-cost prosthetic hand that has reliable pattern recognition performance. Generally, the cost is minimized by reducing the number of channels and the degrees of freedom used [80]. One of the inherent reasons for compromising on the degrees of freedom in a low-cost myoelectric pattern recognition system is the least separable margin among the movements considered, especially among the movements possessing weak signal strengths [32,46]. Therefore, the objective of this work is to obtain improved EMG pattern recognition performance with a minimal number of channels. To reach this goal, nonlinear scaling-based features, the LMAV and NSV, is proposed. The LMAV and NSV enhance the EMG pattern recognition performance when the LMAV and NSV are grouped with each of the existing feature extraction methods considered (FS1, FS2, FS3, and FS4). This performance enhancement indicates that the LMAV and NSV add some new information to the existing feature extraction methods due to the use of nonlinear scaling on the signal amplitude rather than using the original signal amplitude used in [26]. The nonlinear scaling operation yields higher discrimination among weak signals than strong signals and thus contributes to the performance enhancement. In this study, we also propose a combined feature extraction method, including the proposed LMAV and NSV, along with the existing WL, WAMP, SSC, ZC, MOB, COM, SKW, and four autoregressive coefficients, which achieves the highest EMG pattern recognition performance in terms of accuracy, sensitivity, specificity, precision and F1 score compared with the existing feature extraction methods [36,37,60,61]. However, the proposed feature extraction method enhances accuracy, sensitivity, specificity, precision, and F1 score by 1.00%, 5.01%, 0.55%, 4.71%, and 5.06%, respectively. Additionally, on dataset 2 with LDA classifier, the proposed method improves accuracy, sensitivity, specificity, precision, and F1 score by 1.18%, 5.90%, 0.66%, 5.63%, and 6.04%, respectively.

In this study, the LMAV and NSV are validated across two identical EMG datasets (one newly collected in our lab and one standard), where each of the datasets employs two channels and five individual and five combined finger movements. However, these datasets employ distinct acquisition systems, electrodes, processing circuits, numbers of bits for ADC, and sampling frequencies. Again, an interesting finding is that the proposed LMAV and NSV contribute to improved EMG pattern recognition performances for both datasets; this proves the strength of the LMAV and NSV for the standard dataset and the experimental dataset that we have collected from our experimental acquisition system.

Again, the LMAV and NSV yield consistent performance enhancements over window sizes ranging from 50 ms to 350 ms, thereby ensuring the applicability of the LMAV and NSV over various window sizes. In this evaluation, we consider the maximum window size to be 350 ms since a disjoint window size higher than 250 ms does not contribute to significantly enhancing the EMG pattern recognition performance and increases the system delay [65]. In addition, the most noticeable characteristics of the LMAV and NSV are that the proposed features mostly improve the movement-wise F1 score (up to 2.60% and 2.30% for dataset 1 and dataset 2, respectively), and this enhances the strength of the LMAV and NSV.

It is also important to note that the proposed feature extraction method show stable EMG pattern recognition performances across the LDA, SVM, and KNN classifiers. Therefore, the proposed feature extraction method provides an option when choosing a classifier for a given requirement.

Our study has some limitations. Dataset 1 was collected using a wet electrode; so, the noise in no movement condition was very less compared to the EMG signal during muscle contraction (average SNR lies between 5 dB to 23 dB). But, in the case of using a dry electrode, the noise may increase including power line artefacts and AWGN. So, to generate the noisy condition artificially, the EMG signal is contaminated with AWGN using MATLAB R2017a environment [74]. But, at low SNR values, the EMG pattern recognition performance of the proposed feature extraction method is found less in comparison with the performance at high SNR; but everywhere the proposed features (LMAV and NSV) contribute more or less to enhancing the performance. Again, the current EMG pattern recognition performance includes steady-state EMG datasets (dataset 1 and dataset 2) only; but a real-time prosthetic control includes both steady-state and transient condition. So, further study is required considering dry electrodes and transient condition. In addition to these, the contribution of the newly proposed LMAV and NSV in terms of pattern recognition performance will be investigated for a multichannel electrode array. A similar investigation will also be carried out for the proposed feature extraction method. In addition, the performance of the proposed feature extraction method on the other physiological will be investigated.

## 5. Conclusions

Two nonlinear scaling-based features, the LMAV and NSV, are proposed and validated across two datasets for four existing feature extraction methods and three classifiers. The experimental results indicate that the proposed features significantly enhance the EMG pattern recognition performances yielded when they are grouped with the existing feature extraction methods. It is also mentioned that the important strengths of the proposed features are stable performance enhancements on both datasets, with classifiers, under a variable window size, and a variable SNR. In addition to the newly proposed features, we also propose a combined feature extraction method (the LMAV and NSV along with the existing 11 time-domain features), which achieves the best performances on both datasets and with all three classifiers. In this study, FS1 with the LMAV and NSV achieves the second-best EMG pattern recognition performance across all cases. Moreover, it is also noted that the LDA classifier provides better performance than the SVM and KNN.

## Appendix A

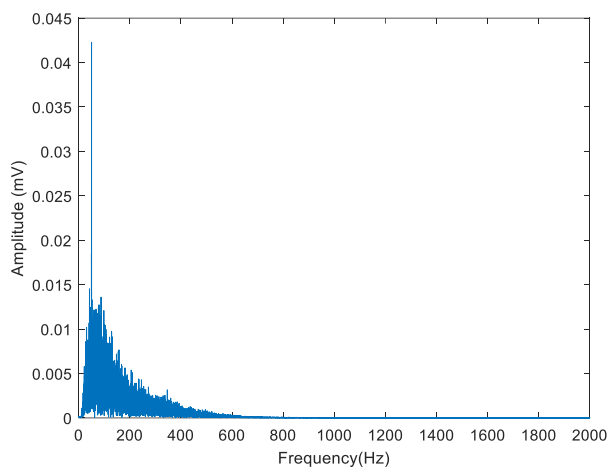

**Figure A.** The frequency spectrum of an EMG signal of dataset 2.

# Appendix B

**Table B1:** The *p*-values between the proposed feature extraction method and each of the existing feature extraction method.

| Parameter | Classifier | FS1 | FS2 | FS3 | FS4 |
|---|---|---|---|---|---|
| Accuracy | LDA | p<0.001 | p<0.001 | p<0.001 | p<0.001 |
|  | SVM | p<0.001 | p<0.001 | p<0.001 | p<0.001 |
|  | KNN | p<0.001 | p<0.001 | p<0.001 | p<0.001 |
| Sensitivity | LDA | p<0.001 | p<0.001 | p<0.001 | p<0.001 |
|  | SVM | p<0.001 | p<0.001 | p<0.001 | p<0.001 |
|  | KNN | p<0.001 | p<0.001 | p<0.001 | p<0.001 |
| Specificity | LDA | p<0.001 | p<0.001 | p<0.001 | p<0.001 |
|  | SVM | p<0.001 | p<0.001 | p<0.001 | p<0.001 |
|  | KNN | p<0.001 | p<0.001 | p<0.001 | p<0.001 |
| Precision | LDA | p<0.001 | p<0.001 | p<0.001 | p<0.001 |
|  | SVM | p<0.001 | p<0.001 | p<0.001 | p<0.001 |
|  | KNN | p<0.001 | p<0.001 | p<0.001 | p<0.001 |
| F1 Score | LDA | p<0.001 | p<0.001 | p<0.001 | p<0.001 |
|  | SVM | p<0.001 | p<0.001 | p<0.001 | p<0.001 |
|  | KNN | p<0.001 | p<0.001 | p<0.001 | p<0.001 |

**Table B2:** The *p*-values between group 1 and group 2.

| Parameter | Classifier | FS1 | FS2 | FS3 | FS4 |
|---|---|---|---|---|---|
| Accuracy | LDA | 0.001 | 0.210 | p<0.001 | p<0.001 |
|  | SVM | p<0.001 | p<0.001 | p<0.001 | p<0.001 |
|  | KNN | p<0.001 | p<0.001 | p<0.001 | p<0.001 |
| Sensitivity | LDA | 0.001 | 0.194 | p<0.001 | p<0.001 |
|  | SVM | p<0.001 | p<0.001 | p<0.001 | p<0.001 |
|  | KNN | p<0.001 | p<0.001 | p<0.001 | p<0.001 |
| Specificity | LDA | 0.002 | 0.199 | p<0.001 | p<0.001 |
|  | SVM | p<0.001 | p<0.001 | p<0.001 | p<0.001 |
|  | KNN | p<0.001 | p<0.001 | p<0.001 | p<0.001 |
| Precision | LDA | 0.001 | 0.069 | p<0.001 | p<0.001 |
|  | SVM | p<0.001 | p<0.001 | p<0.001 | p<0.001 |
|  | KNN | p<0.001 | p<0.001 | p<0.001 | p<0.001 |
| F1 Score | LDA | 0.002 | 0.198 | p<0.001 | p<0.001 |
|  | SVM | p<0.001 | p<0.001 | p<0.001 | p<0.001 |
|  | KNN | p<0.001 | p<0.001 | p<0.001 | p<0.001 |

**Table B3:** The *p*-values between the F1 scores of 11-features and 11-features along with LMAV and NSV at various window size.

| Window size (ms) | *p*-value |
|---|---|
| 50 | p<0.001 |
| 100 | p<0.001 |
| 150 | p<0.001 |
| 200 | p<0.001 |
| 250 | p<0.001 |
| 300 | p<0.001 |
| 350 | p<0.001 |

**Table B4:** The *p*-values between the F1 scores of 11-features and 11-features along with LMAV and NSV at various SNR.

| SNR (dB) | *p*-value | SNR (dB) | *p*-value | SNR (dB) | *p*-value |
|---|---|---|---|---|---|
| 0 | 0.265 | 7 | 0.008 | 14 | p<0.001 |
| 1 | 0.163 | 8 | 0.019 | 15 | 0.132 |
| 2 | 0.118 | 9 | 0.079 | 16 | 0.044 |
| 3 | 0.039 | 10 | 0.071 | 17 | 0.002 |
| 4 | 0.019 | 11 | 0.019 | 18 | 0.010 |
| 5 | 0.006 | 12 | 0.044 | 19 | p<0.001 |
| 6 | 0.700 | 13 | 0.083 | 20 | 0.010 |

## Conflicts of Interest

All authors declare that there are no conflicts of interest regarding the publication of this paper.

## Acknowledgements

The authors would like to show their sincere gratitude towards all the participants of EMG dataset 1, and Dr. Rami N. Khushaba for making EMG dataset 2 publicly available on his website. The website for the database is https://www.rami-khushaba.com/electromyogram-emg-repository.html.

## Funding

This research is financially supported by Xiamen University Malaysia, Project number XMUMRF/2018-C2/IECE/0002, the Information and Communication Technology Division, Ministry of Posts, Telecommunications and Information Technology, Government of Bangladesh under reference number 56.00.0000.028.33.098.18-219 and Universiti Kebangsaan Malaysia under grant number DPK-2021-001, GUP-2021-019, and TAP-K017701.